\documentclass[twocolumn,showpacs,preprintnumbers,amsmath,amssymb,prb,superscriptaddress]{revtex4} 
 
\usepackage{epsfig}
\usepackage[dvips]{color}
\usepackage{graphicx}
\usepackage{dcolumn}
\usepackage{bm}
 
\begin{document} 
 
\title{Complete bond--operator theory of the two--chain spin ladder}
 
\author{B. Normand} 
\affiliation{Department of Physics, Renmin University of China, Zhongguancun 
Ave. 59, Beijing 100872, China} 
 
\author{Ch.~R\"uegg}
\affiliation{London Centre for Nanotechnology and Department of Physics and 
Astronomy, University College London, London WC1E 6BT, United Kingdom}

\date{\today} 

\begin{abstract} 

The discovery of the almost ideal, two--chain spin--ladder material 
(C$_5$H$_{12}$N)$_2$CuBr$_4$ has once again focused attention on this 
most fundamental problem in low--dimensional quantum magnetism. Within 
the bond--operator framework, three qualitative advances are introduced 
which extend the theory to all finite temperatures and magnetic fields 
in the gapped regime. This systematic description permits quantitative 
and parameter--free experimental comparisons, which are presented for 
the specific heat, and predictions for thermal renormalization of the 
triplet magnon excitations.

\end{abstract} 
 
 
\maketitle 

\section{Introduction} 
 
Quantum magnetic systems have in recent years offered a wealth of 
opportunities for the study of a broad range of novel physical phenomena. 
Systematic improvements in the growth of pure, single--crystalline samples 
of low--dimensional spin systems, and in the measurement techniques applied 
to probe their microscopic properties, have resulted in a remarkable 
convergence of theory and experiment to explain a number of fundamental 
observations. At the foundation of many of these breakthroughs has been 
the two--chain ``spin--ladder'' geometry of stacked, interacting dimer 
units, shown in Fig.~1. First synthesized in cuprate systems,\cite{rhatb} 
both undoped (spin) and doped ladders have been the subject of extensive 
theoretical investigation in the context of gapped and gapless quantum 
magnetism,\cite{rdr} resonating valence--bond states,\cite{rwns} generalized 
frustration effects,\cite{roos,rgnb} impurity doping,\cite{rsf} (charge) 
Luttinger--liquid phases,\cite{rhp,rlbf} and superconductivity.\cite{rdrs,rsrz}

Focusing on dimer--based quantum spin systems, examples of physical 
phenomena realized rather well in experiment include magnetic quantum 
phase transitions\cite{rgrt,rrfsskgm} and quantum critical regimes,\cite{rgrt} 
unconventional, critical longitudinal magnons,\cite{rrnmfmkggmb} the 
Bose--Einstein condensation of magnetic excitations,\cite{rnoot,rrn} and 
the demonstration of their statistical properties.\cite{rttw,rrnmnfkgbsm} 
However, none of this rich physics is intrinsically a consequence of 
one--dimensionality: while experimental efforts to create ladder--like 
geometries with a range of coupling scales and ratios have been rather 
successful, until recently none of the candidate materials has been very 
accurately one--dimensional (1D) in nature. This situation has now been 
changed by the organometallic compound (C$_5$H$_{12}$N)$_2$CuBr$_4$, a 
system of two--chain spin ladders whose net interladder coupling is 
approximately 3\% of the characteristic ladder energy scale, which 
propels this geometry back to the forefront of exploration into the 
unique properties of 1D magnets. 

\begin{figure}[t]
\includegraphics[width=8cm]{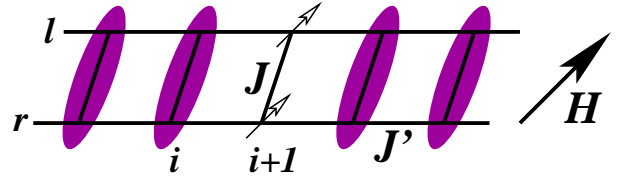}
\caption{(color online) Schematic representation of the spin ladder. 
$S = 1/2$ spins at each lattice site are taken in the bond--operator 
representation to be correlated predominantly with the spin on the same 
ladder rung, forming a singlet state (shaded ellipse), except when the 
rung contains a propagating triplet state (depicted as separate spins). 
$J$ is the exchange interaction on a dimer bond, $J'$ the interaction 
on the chain bonds, and $l$ and $r$ are indices for the two chains. 
The field direction sets the triplet quantization axis.}
\label{lf1}
\end{figure}

(C$_5$H$_{12}$N)$_2$CuBr$_4$\cite{rpsw,rwea} has the additional, major 
advantages of forming large, readily deuterated, single crystals and of 
its energy scales falling within the range of laboratory magnetic fields.
The magnetic field is an essential parameter in both experimental 
and theoretical studies. The observation, made at a rather early stage, 
that a spin ladder in a field should be driven into a quantum critical, 
spin Luttinger--liquid regime,\cite{rsss} was followed by a number of 
numerical\cite{rhpl} and theoretical studies\cite{rcg,rm} of the ground 
state and low--energy excitations. However, the experimental situation 
now mandates a physical description which captures all of the key quantum 
fluctuation effects in a spin ladder at finite temperatures and applied 
fields. 

The impressive arsenal of experimental techniques available to modern 
condensed matter physics has in a very short time revealed from the 
(C$_5$H$_{12}$N)$_2$CuBr$_4$ system alone not only most of the phenomena 
listed in the second paragraph but also the field--induced spin Luttinger 
liquid. The latter is a direct consequence of the impressively 1D nature 
of the material. Thermodynamic measurements, specifically of thermal 
expansion,\cite{rlea} specific heat,\cite{rcrct} and the magnetocaloric 
effect\cite{rcrct,rbt3d} have elucidated the phase diagram: the quantum 
disordered phase gives way to the Luttinger--liquid regime at $H_c = 6.99$ 
T, full saturation is achieved at $H_s = 14.4$ T, and Luttinger--liquid
behavior can be followed up to temperatures of order 1.5 K at $H \sim 11$ T. 
Nuclear magnetic resonance (NMR)\cite{rkea} has been used to measure critical 
and Luttinger--liquid exponents, the latter varying with the applied field. 
The field--induced 3D ordered phase, which is a consequence of the weak 
interladder coupling and sets in below 100 mK, has been found by 
NMR\cite{rkea} and probed in detail by neutron diffraction.\cite{rbt3d} 
Fractionalization of the characteristic ladder magnon excitations into 
field--controlled spinon continua, which constitute unambiguous proof for 
a Luttinger--liquid description over a wide range of energy scales in the 
lowest triplet branch, has been demonstrated by inelastic neutron scattering 
(INS).\cite{rbtins}

The spin ladder is the quintessential gapped 1D quantum magnet: with its 
short correlation length, it is perfectly suited for numerical approaches, 
and many of its properties have been computed with great accuracy by exact 
diagonalization (ED)\cite{rbdrs} and density--matrix renormalization--group 
(DMRG)\cite{rwns} calculations of the ground and excited states, augmented 
by quantum Monte Carlo (QMC)\cite{rttw} and transfer--matrix 
renormalization--group (TMRG)\cite{rwy} studies of thermodynamics. It 
has also been used in the development of systematic, high--order techniques 
such as series expansions\cite{rtmhzs} and Continuous Unitary Transformations 
(CUTs)\cite{rksgu} which give quantitatively accurate information concerning 
high--energy excitations, including bound states and spectral weights. 
Simultaneously, however, the spin ladder is rather poorly described by most 
analytical approaches, as much of the sophisticated machinery available for 
1D systems is applicable to gapless quantum magnets. Indeed, even the 
numerical approaches listed are all incomplete in some regard, most notably 
concerning finite--temperature dynamics.

The purpose of this manuscript is to develop one straightforward analytical 
description, based on the bond--operator representation of the dimer spin 
states, to give a complete account of the static and dynamic properties of 
the two--chain spin ladder in the spin--gap regime. The development requires 
three qualitative additions to the basic bond--operator formalism, 
specifically finite interdimer correlation terms, their evolution with 
field, and the incorporation of effective statistics for magnon excitations. 
Given the exchange interactions of the ladder, the description has no free 
parameters. 

The focus here will be on the low--field, quantum disordered phase of the 
ladder, where the spin gap is a consequence of quantum fluctuations. The 
high--field, saturated regime of the ladder is also gapped, and is also 
separated from the Luttinger--liquid phase by a crossover (which in higher 
dimensions is a true quantum phase transition), but here the quantum 
fluctuation effects are precisely zero and the gap is a consequence only 
of the field. The zero--temperature properties of this regime were described 
in Ref.~[\onlinecite{rn}]. Thermal fluctuation effects do, however, mandate 
the same treatment of thermodynamics and finite--temperature dynamics as that 
presented here for the quantum disordered phase.  

The structure of this paper is as follows. Section II presents the extended 
bond--operator formalism. In Sec.~III this is applied at zero temperature to 
review the basic properties and fundamental physics of the spin ladder as a 
function of the ratio of chain to rung interaction parameters. In Sec.~IV, 
the effective statistical properties of the magnon excitations are introduced, 
and employed to solve the mean--field equations of the ladder at finite 
temperatures and magnetic fields. Illustrative results for the isotropic 
ladder are used to highlight the physics captured by the complete 
bond--operator description. In Sec.~V, quantitative calculations are 
performed for the parameters of (C$_5$H$_{12}$N)$_2$CuBr$_4$, and 
particular attention is paid to a comparison with experimental data 
for the specific heat. A summary is presented in Sec.~VI. 

\section{Bond--operator mean--field theory}

The Hamiltonian for a Heisenberg spin ladder in an external magnetic field 
(Fig.~1) is 
\begin{equation} 
H = \sum_{i} \! J {\vec S}_{i}^l {\bf \cdot}{\vec S}_{i}^r + \!\! 
\sum_{i,m=l,r} \!\! J' {\vec S}_{i}^m {\bf \cdot} {\vec S}_{i+1}^m
 + \sum_{i} {\vec h}_i {\bf \cdot} ({\vec S}_i^l + {\vec S}_i^r) 
\label{essh} 
\end{equation} 
with ${\vec h} = g \mu_{\rm B} {\vec H}$. Here one considers a minimal, 
purely 1D model for SU(2) spins with no additional couplings (such as 
frustrating, next--neighbor, three-- or four--spin interactions) and 
no anisotropies (which in real materials could be of exchange, 
Dyzaloshinskii--Moriya, single--ion, $g$--tensor, or other origin). The 
Hamiltonian is transformed using the identity 
\begin{equation} 
S_{\alpha}^{l,r} = \pm {\textstyle \frac{1}{2}} (s^{\dag}  
t_{\alpha} + t_{\alpha}^{\dag} s) - i \epsilon_{\alpha \beta \gamma}  
t_{\beta}^{\dag} t_{\gamma},
\label{ebot} 
\end{equation} 
where $s_i$ and $t_{i,\alpha} (\alpha = x,y,z$) are bond operators for the 
singlet and triplet states of each ladder rung.\cite{rsb,rgrs} These operators 
have bosonic statistics, required to reproduce the spin algebra of 
$S_{\alpha}^{l,r}$, but they must also obey the local hard--core constraint, 
\begin{equation} 
s_{i}^{\dag} s_{i} + \sum_{\alpha} t_{i,\alpha}^{\dag} t_{i,\alpha} = 1, 
\label{eboc} 
\end{equation} 
on each ladder rung. This is the physical statement that each rung may 
only be in a singlet state or one of the three triplets (equivalent to 
the four possible states of two spin--1/2 entities): these operators 
are hard--core bosons. In the presence of a magnetic field, and the 
absence of spin--space anisotropy, it is convenient to transform to the 
basis $\alpha = +,0,-$.\cite{rmnrs} 

In the bond--operator representation, the minimal Hamiltonian 
of Eq.~(\ref{essh}) takes the form $H = H_0 + H_2 + H_4$, 
where\cite{rgrs,rnr1,rnr2,rmnrs} 
\begin{eqnarray} 
H_0 & \! = \! & \sum_{i} - J ({\textstyle \frac{3}{4}} s_{i}^{\dag}  
s_{i} \! + \! {\textstyle \frac{1}{4}} t_{i,\alpha}^{\dag} t_{i,\alpha})  
\! - \! \mu_i ( s_{i}^{\dag} s_i \! + \! t_{i,\alpha}^{\dag} 
t_{i,\alpha} \! - \! 1) \nonumber \\ & & + \sum_i h (t_{i+}^{\dag} t_{i,+}
 - t_{i,-}^{\dag} t_{i-}), 
\label{ebosh0} 
\end{eqnarray} 
with summation over the repeated index $\alpha$,
\begin{equation} 
H_2 = {\textstyle \frac{1}{2}} J' \sum_{i,\alpha} t_{i,\alpha}^{\dag} 
t_{i+1, \alpha} s_{i+1}^{\dag} s_i + t_{i,\alpha}^{\dag} t_{i+1,
{\overline{\alpha}}}^{\dag} s_i s_{i+1} + {\rm H. c.}, 
\label{ebosh2} 
\end{equation} 
with $\overline{\alpha} = -,0,+$, and 
\begin{eqnarray} 
H_4 & = & {\textstyle \frac{1}{4}} J' \sum_{i} [ t_{i,0}^{\dag} t_{i+1,0}
(t_{i,+}^{\dag} t_{i+1,+} + t_{i,-}^{\dag} t_{i+1,-}) \\ & & 
- t_{i,0} t_{i+1,0} (t_{i,+}^{\dag} t_{i+1,-}^{\dag} + t_{i,-}^{\dag} 
t_{i+1,+}^{\dag}) + {\rm H. c.}] \nonumber \\ & & + (t_{i,+}^{\dag} t_{i,+}
\! - \! t_{i,-}^{\dag} t_{i,-}) (t_{i+1,+}^{\dag} t_{i+1,+} \! - \! 
t_{i+1,-}^{\dag} t_{i+1,-}) . \nonumber
\label{ebosh4} 
\end{eqnarray} 
The second term in $H_0$ enforces the constraint (\ref{eboc}) using the 
Lagrange multipliers $\mu_i$. At zero magnetic field, the term quadratic 
in the singlet operators is negative, ensuring a singlet condensation and 
justifying the replacement $s_i = \langle s_i \rangle$ on each dimer. The 
ground state of the system is then described as a condensate of singlets 
with a spin gap to all triplet excitations. For a uniform ladder there 
is no site variation, and hence one may take $\langle s_i \rangle = 
\overline{s}$ and $\mu_i = \mu$ (at which point it is clear that the 
constraint is enforced only globally, rather than locally). In the absence 
of terms violating the inversion symmetry of the rung bonds, there is no 
three--triplet contribution. The four--triplet interaction term $H_4$ is 
written explicitly to facilitate the mean--field decomposition to follow. 

In the gapped phase of the spin ladder, the triplet operators obey $\langle 
t_{i,\alpha} \rangle = 0$, but there is no such condition on two--dimer 
correlation functions. To include the finite--field case, one may generalize 
immediately by introducing the three quantities $P_\alpha = \langle 
t_{i,\alpha}^{\dag} t_{i+1,\alpha} \rangle$ and the pair $Q_0 = \langle 
t_{i,0}^{\dag} t_{i+1,0}^{\dag} \rangle$, $Q_{+-} = \langle t_{i,+}^{\dag} 
t_{i+1,-}^{\dag} \rangle = \langle t_{i,-}^{\dag} t_{i+1,+}^{\dag} \rangle 
 = Q_{-+} \equiv Q$; in the $+,0,-$ basis, all five expectation values are 
real. First introduced in Ref.~[\onlinecite{rgrs}] for the zero--field case, 
the quantities $P$ and $Q$ encode respectively the quantum fluctuations for 
triplet hopping along the ladder and for spontaneous creation of triplet 
pairs on neighboring ladder rung bonds. In terms of the spins at each site, 
$P$ and $Q$ describe the nearest--neighbor spin correlation functions along 
the ladder direction, which will be the subject of Sec.~IIIB. Further, at 
finite temperature, a finite magnetic field will permit a non--zero 
magnetization $m_i = \langle t_{i,+}^{\dag} t_{i,+} \rangle - \langle 
t_{i,-}^{\dag} t_{i,-} \rangle$, which will be uniform ($m_i = m$) and 
whose inclusion constitutes the third and final qualitative extension of 
the basic framework. 

When the Hamiltonian is decoupled using the finite expectation values of 
the preceding paragraph, transformed to reciprocal space, and diagonalized 
in the conventional manner,\cite{rgrs,rnr1,rnr2,rmnrs} the net result is the 
quadratic triplet Hamiltonian 
\begin{equation} 
H_{\rm mf} = e_0 + \sum_{k,\alpha} \omega_{k,\alpha} \gamma_{k,\alpha}^{\dag} 
\gamma_{k,\alpha} \left( \textstyle{\frac{1}{2}} + n (\omega_{k,\alpha}) 
\right),
\label{ehmf}
\end{equation}
where the sums for each branch $\alpha$ are normalized to the system size 
and the zeroth--order contribution is 
\begin{eqnarray} 
e_0 & = & (- \textstyle{\frac{3}{4}} J - \mu) \overline{s}^2 
+ \textstyle{\frac{5}{2}} \mu - \textstyle{\frac{3}{8}} J 
 - \textstyle{\frac{1}{2}} J' m^2 \\ & & 
 - \textstyle{\frac{1}{2}} J' [ P_+^2 + P_-^2 + 2 P_0 (P_+ + P_-)
 - 2 Q^2 - 4 Q Q_0] \nonumber
\label{eeo}
\end{eqnarray}
per dimer. The operator $\gamma_{k,\alpha}$ is the bosonic quasiparticle 
diagonalizing the Hamiltonian matrix and the eigenenergies are 
\begin{eqnarray}
\omega_{k,+} & = & \omega_k + \textstyle{\frac{1}{2}} J' \cos k \, (P_+ - 
P_-) + J' m - h, \label{ewp} \\
\omega_{k,0} & = & \sqrt{\Lambda_{k,0}^2 - \Delta_{k,0}^2},
\label{ewo} \\
\omega_{k,-} & = & \omega_k - \textstyle{\frac{1}{2}} J' \cos k \, (P_+ - 
P_-) - J' m + h, \label{ewm}
\end{eqnarray}
with 
\begin{eqnarray}
\omega_k & = & \sqrt{\Lambda_k^2 - \Delta_k^2}, \label{ew} \\ 
\Lambda_k & = & \textstyle{\frac{1}{4}} J - \mu + J' \cos k \, 
(\overline{s}^2 + P_0 + \textstyle{\frac{1}{2}} (P_+ + P_-)), \nonumber \\
\Delta_k & = & J' \cos k \, (\overline{s}^2 - Q - Q_0), \nonumber \\
\Lambda_{k,0} & = & \textstyle{\frac{1}{4}} J - \mu + J' \cos k \,
(\overline{s}^2 + P_+ + P_-), \nonumber\\
\Delta_{k,0} & = & J' \cos k \, (\overline{s}^2 - 2 Q). \nonumber
\end{eqnarray}
The quantity $n (\omega_{k,\alpha})$, which enters at finite temperatures, 
is the thermal occupation beyond the zero--point contribution. In a fully 
rigorous treatment it is the Bose function. Here, however, it will be 
replaced by an effective thermal occupation function which captures 
correctly the hard--core nature of the magnons at high temperatures. A 
detailed discussion is deferred to Sec.~IV.

The mean--field equations obtained by taking the derivatives of $H_{\rm mf}$ 
(\ref{ehmf}) with respect to $\mu$ and $\overline{s}$ are 
\begin{eqnarray} 
\textstyle{\frac{5}{2}} \! - \! \overline{s}^2 & \!\! = \!\! & \sum_k 
\frac{\Lambda_k}{\omega_k} \tilde{n} (\omega_{k,+} \! ) \! + \! 
\frac{\Lambda_{k,0}}{\omega_{k,0}} \tilde{n} (\omega_{k,0} \! ) 
\! + \! \frac{\Lambda_k}{\omega_k} \tilde{n} (\omega_{k,-} \! ), 
\label{emfem}
\\ \textstyle{\frac{3}{4}} J \! + \! \mu & \!\! = \!\! & \sum_k J' \cos k 
\left[ \frac{\Lambda_k \! - \! \Delta_k}{\omega_k} \tilde{n} (\omega_{k,+}) 
\right. \\ & & \;\;\;\;\; \left. + \frac{\Lambda_{k,0} \! - \! 
\Delta_{k,0}}{\omega_k} \tilde{n} (\omega_{k,0}) + \frac{\Lambda_k \! - \! 
\Delta_k}{\omega_k} \tilde{n} (\omega_{k,-}) \right] \!\! , \nonumber
\label{emfes}
\end{eqnarray}
and to the six finite expectation values are 
\begin{eqnarray}
P_0 & \!\! = \!\! & \sum_k \cos k \, \frac{\Lambda_{k,0}}{\omega_{k,0}} 
\tilde{n} (\omega_{k,0}), \\
\label{emfepo}
P_+ & \!\! = \!\! & \sum_k \! \frac{\cos k}{2} \!\! \left[ \! \left( \! 
\frac{\Lambda_k}{\omega_k} \!\! + \!\! 1 \!\! \right) \! \tilde{n} 
(\omega_{k,+} \! ) \! + \!\! \left( \! \frac{\Lambda_k}{\omega_k} \!\!
 - \!\! 1 \!\! \right) \! \tilde{n} (\omega_{k,-} \! ) \! \right] \!\!, \\
\label{emfepp}
P_- & \!\! = \!\! & \sum_k \! \frac{\cos k}{2} \!\! \left[ \! \left( \! 
\frac{\Lambda_k}{\omega_k} \!\! - \!\! 1 \!\! \right) \! \tilde{n} 
(\omega_{k,+} \! ) \! + \!\! \left( \! \frac{\Lambda_k}{\omega_k} \!\!
 + \!\! 1 \!\! \right) \! \tilde{n} (\omega_{k,-} \! ) \! \right] \!\!, \\
\label{emfepm}
Q_0 & \!\! = \!\! & - \sum_k \cos k \, \frac{\Delta_{k,0}}{\omega_{k,0}} 
\tilde{n} (\omega_{k,0}), \\
\label{emfeqo}
Q & \!\! = \!\! & - \sum_k \frac{\cos k}{2} \left[ \frac{\Delta_k}
{\omega_k} \tilde{n} (\omega_{k,+}) + \frac{\Delta_k}{\omega_k} \tilde{n} 
(\omega_{k,-}) \right] \!\!, \\
\label{emfeq}
m & \!\! = \!\! & \sum_k \tilde{n} (\omega_{k,+}) - \tilde{n} 
(\omega_{k,-}), 
\label{emfemm}
\end{eqnarray} 
in which $\tilde{n} (\omega) = \textstyle{\frac{1}{2}} + n (\omega)$. Given 
the overall energy scale $J$, the self--consistent solution of these equations 
at each value of the field $h$ and the temperature $T$, which appears only in 
the thermal occupation function, depends solely on the parameter ratio $J'/J$. 
 
\begin{figure}[t]
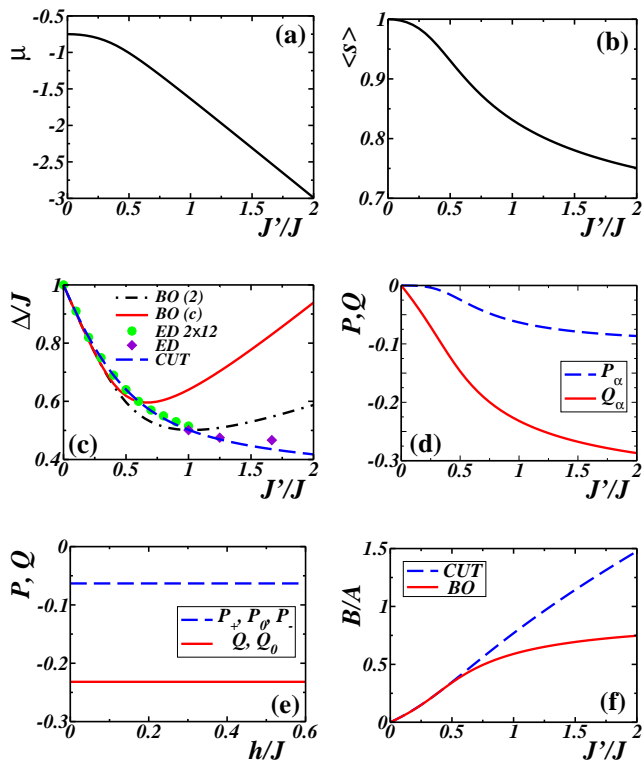

\includegraphics[width=4.1cm]{lm.eps}\hspace{0.3cm}\includegraphics[width=4.0cm]{ls.eps} \\
\phantom{.} 
\vspace{.1cm}
\includegraphics[width=4.1cm]{lgd.eps}\hspace{0.3cm}\includegraphics[width=4.0cm]{lpqlb.eps} \\
\phantom{.}
\vspace{.1cm}
\includegraphics[width=4.1cm]{lpqtob.eps}\hspace{0.3cm}\includegraphics[width=4.0cm]{lscc.eps}
\caption{(color online) Properties of the two--chain ladder as a function 
of coupling ratio $J'/J$ at $h$ = 0 = $T$. (a) Chemical potential $\mu$. (b) 
Singlet condensation fraction $\overline{s} \equiv \langle s \rangle$. (c) 
Spin gap $\Delta$: bond--operator data from Eqs.~(\ref{emfem})--(\ref{emfemm}) 
($BO (c)$, red, solid line); bond--operator data without $P, Q, m$ ($BO (2)$, 
black, dot--dashed line); ED data for 2$\times$12 system ($ED \; 2 \! \times 
\! 12$, green circles); ED data extrapolated to an infinite ladder ($ED$, 
violet diamonds); data from Continuous Unitary Transformations ($CUT$, blue, 
dashed line). (d) Triplet correlations $P$ and $Q$. (e) $P$ and $Q$ for 
$J'/J = 1$, shown as functions of field at $T = 0$. (f) Spin correlation 
ratio $B/A$ (see text), comparing bond--operator and CUT results.}
\label{lf2}
\end{figure}

\section{Zero--temperature ladder physics}

\subsection{Bond--operator description}

An analysis of the physical content of Eqs.~(\ref{emfem})--(\ref{emfemm}) 
begins by considering the qualitative description of the spin ladder at 
zero temperature and zero field. The parameter governing the extent of 
quantum fluctuations away from the limit of decoupled dimers, in which 
the bond--operator approach is exact, but trivial, is the ratio $J'/J$ 
between the exchange interactions on the chain and rung bonds. The essential 
picture is a condensate of rung singlets, which are separated by an energy 
gap of $J$ from the (local) rung triplet states. These form a condensate 
in the true sense of the word, its coherence mediated by the hopping 
processes of virtual triplet excitations (the quantum fluctuations) along 
the ladder.\cite{rsb} The effective chemical potential $\mu$ (Sec.~II) is 
the parameter fixing the band center of the triplet magnon modes, and the 
singlet condensate fraction $\overline{s}$ appears also in the effective 
band width. 

Figure 2(a) shows the evolution of $\mu$ as a function of $J'/J$; because 
the dominant contributions to the integrals in the mean--field equations 
(\ref{emfem})--(\ref{emfemm}) are given by those $k$ values close to $\pi$, 
increasingly negative values of $\mu$ are required to maintain a suitable 
gap $\Delta_\alpha = \omega_{\pi,\alpha}$ as $J'$ is raised (at $h = 0$, 
the three modes $\alpha$ are degenerate). The singlet condensation 
fraction $\overline{s}$ [Fig.~2(b)] decreases from precisely unity in the 
decoupled limit to values of order 80\% beyond the isotropic point ($J'/J
 = 1$), suggesting a rather high level of singlet condensation even far 
from the strongly dimerized limit. The gap of the spin ladder is shown 
in Fig.~2(c), where the bond--operator technique in its simplest, 
``second--order'' form\cite{rgrs} is compared with its complete form, 
and also with data obtained by other techniques. ED data for a 2$\times$12 
ladder was taken from Ref.~[\onlinecite{rrtr}] and that for the extrapolation 
to an infinite ladder from Ref.~[\onlinecite{rbdrs}]. The CUT technique, 
while perturbative, affords a systematic means of proceeding to very high 
order,\cite{rksgu} and can be seen to provide extremely accurate results 
over a wide range of $J'/J$ for a gapped system ({\it i.e.}~one with a 
short correlation length) such as the two--chain ladder. 

Within the bond--operator description, large quantitative errors are 
inevitable at $J' > J$, where the framework loses its validity: the 
gap begins to rise again as a consequence of the band center increasing 
while $\overline{s}$ falls with $J'$ [Fig.~2(b)]. Although the gap prediction 
is accurate only at small $J'/J$, the second--order approximation delivers 
serendipitously good agreement with the exact result precisely at the 
isotropic point, where $\Delta_{\rm BO}^{(2)} (1) = 0.503 J$ while $\Delta
_{\rm ED}^\infty (1) = 0.501 J$. Although the inclusion of the expectation 
values $P$ and $Q$ was addressed in Ref.~[\onlinecite{rgrs}], a missing 
factor of 3 for the triplet mode degeneracy caused errors in the results 
for all quantities. In combination with the above numerical coincidence, this 
led to rather little attention being paid to the $P$ and $Q$ terms in the 
literature. Their qualitative importance will become clear in Secs.~IV and 
V. Quantitatively, in fact the complete bond operator result at the isotropic 
point is $\Delta_{\rm BO}^{(c)} (1) = 0.640 J$, a value which is close to 
half--way between the second-- and third--order approximations to the 
gap obtained in a strong--dimerization expansion.\cite{rrtr} Thus one may 
conclude that the results obtained from Eqs.~(\ref{emfem})--(\ref{emfemm}) 
are essentially optimal within the bond--operator framework, and that this 
is by its nature limited to be quantitatively accurate only up to $J'/J 
\simeq 0.5$. The qualitative physics of the spin ladder may, however, be 
captured for considerably higher values of the coupling ratio. 

At finite fields but zero temperature, the three magnon bands are split by 
the Zeeman effect, the uniform linear decrease of the entire $+$ branch and 
increase of the $-$ branch with a slope determined by the $S = 1$ nature of 
the magnons and by the $T = 0$ $g$--factor. The zero--field gap is an 
important quantity because it determines the critical field required to 
render the lowest mode gapless, and thus to drive the system into the 
Luttinger--liquid regime at the lowest temperatures. 

\subsection{Spin correlations}

Turning to the interdimer correlations $P$ and $Q$ at zero field 
and temperature, the evolution of the two quantities as a function of 
$J'/J$ is shown in Fig.~2(d). Their negative sign is a consequence of 
the relative phase of the triplets. While the triplet hopping term $P$ 
obviously remains rather small for all ratios, the pair--creation 
term $Q$ constitutes a very significant quantum fluctuation contribution 
even at moderate values of $J'/J$. For perspective, in the isotropic ladder 
($J'/J = 1$) one finds that each of the three components [$Q$ appears twice 
(\ref{ew})] has a value of order 0.23; given that $\overline{s} = 0.83$ 
for $J'/J = 1$ [Fig.~2(b)], it is largely the $Q$ terms which exhaust the 
relevant sum rules. The dominance of the triplet pair--creation term found 
here is fully consistent with studies of the $t$--$J$ ladder.\cite{rsrz} 

Figure 2(e) illustrates the initially non--intuitive result that the 
zero--temperature triplet correlation functions do not vary at all with 
the applied magnetic field: $P_+ = P_0 = P_-$ and $Q = Q_0$ at all fields, 
reflecting the unbroken SU(2) spin symmetry of the electronic wave function 
in the gapped phase. However, it becomes clear by inspection of the 
mean--field equations (\ref{emfem})--(\ref{emfemm}) that in fact a finite 
temperature is required before the applied field can affect either the $P$ 
and $Q$ terms or the magnetization $m$. This behavior is the topic of 
Sec.~IV. 

There is a direct relationship between the quantities $P$ and $Q$ and 
the interdimer spin correlations $\langle {\vec S}_i^l {\bf \cdot} {\vec 
S}_{i+1}^l \rangle = \langle {\vec S}_i^r {\bf \cdot} {\vec S}_{i+1}^r 
\rangle \equiv B$: from Eq.~(\ref{ebot}) it follows that, up to 
fourth--order corrections, $B = \textstyle{\frac{3}{2}} \overline{s}^2 
(P + Q)$. Further, the on--dimer spin correlation $\langle {\vec S}_i^l 
{\bf \cdot} {\vec S}_{i}^r \rangle \equiv A$ is straightforwardly $A =
 - \textstyle{\frac{3}{4}} \overline{s}^2$, which clearly has the required 
limit as $J'/J \rightarrow 0$ and $\overline{s} \rightarrow 1$, and hence 
the ratio $B/A = - 2 (P + Q)$. This ratio has been measured\cite{rsgea,rtu} 
by neutron--scattering studies of (C$_5$H$_{12}$N)$_2$CuBr$_4$, and the 
results are consistent with the ratio $J'/J \simeq 1/3.9$ deduced for 
this material (Sec.~V). Bond--operator results for $B/A$ are compared 
in Fig.~2(f) with the CUT analysis of the spin ladder, also presented 
in Ref.~[\onlinecite{rsgea}], over a range of coupling ratios $J'/J$. As 
for the gap, the bond--operator technique delivers quantitative accuracy 
up to $J'/J \simeq 0.5$, while deviations appear and increase beyond this. 

Thus the intrinsic nature of the quantum mechanical wave function of the 
spin ladder, by which is meant the extent of spin correlations across the 
different bonds, may be deduced directly from the triplet correlations.
Their evolution due to quantum fluctuations (this Sec.) and to thermal 
fluctuations (Sec.~IV), as well as the combined effects of both types 
of fluctuation in a magnetic field, may be followed in the bond--operator 
approach. This type of analysis may be applied to a number of gapped, 
low--dimensional quantum spin systems, including the dimerized $S = 1/2$ 
chain Cu(NO$_3$)$_2 \cdot 2.5$D$_2$O\cite{rxbra} and the 2D $S = 1/2$ 
system PHCC,\cite{rszrb} for which systematic measurements of the 
quantities $J_{ij} \langle {\vec S}_i \cdot {\vec S}_{j} \rangle$ 
have already been performed.

\section{Finite temperatures}

Turning now to the situation at finite temperature, there is naturally 
no explicit expression for a thermal occupation function of hard--core 
bosons. This problem was addressed for the ladder system in 
Ref.~[\onlinecite{rttw}] by enforcing the local constraint (\ref{eboc}) 
at the global level to obtain an effective partition function for the 
hard--core boson system. The effective statistics deduced from this 
expression, 
\begin{equation}
n(\omega_{k,\alpha},\beta) = \frac{e^{-\beta \omega_{k,\alpha}}}{1 + 
\sum_\alpha z_\alpha (\beta)},
\label{eehcbs}
\end{equation}
where $\beta$ is the inverse temperature and $z_\alpha (\beta) = \sum_k 
e^{-\beta \omega_{k,\alpha}}$ is the partition function of mode $\alpha$, 
were shown in Ref.~[\onlinecite{rrnmnfkgbsm}] to give a spectacularly 
successful description of the thermal band--narrowing and gap enhancement 
measured in the 3D coupled--dimer system TlCuCl$_3$. However, it should 
be emphasized that the global reweighting of the partition function 
formulated for a spin ladder in Ref.~[\onlinecite{rttw}] is entirely 
dimension--independent, and thus should function equally well for all 
hard--core boson systems. Indeed the same ansatz has also been used to 
obtain good descriptions of thermodynamic quantities in the 2D dimerized 
spin--chain system (VO)$_2$P$_2$O$_7$\cite{run} and of spectral weights
for the excitations of ultra--cold bosonic atoms on a 1D optical 
lattice.\cite{rrsu}

It is important to recall the meaning of Eqs.~(\ref{emfem})--(\ref{emfemm}) 
at finite temperatures. The spectrum of excitations of any Hamiltonian is 
a temperature--independent quantity, and the temperature is responsible 
only for driving changes in the weights with which given modes are occupied. 
These spectral weights are represented by the dynamic structure factor 
$S({\vec k},\omega, T)$, which can be measured by INS. The bond--operator 
approach delivers an effective description of the one--magnon sector of the 
dynamical structure factor, in the form of a peak position for each of the 
(field--split) magnon branches. Changes in these peak positions as a 
function of temperature are in fact a statement of the leading thermal 
effects on the spectral--weight distribution: the key result of 
Ref.~[\onlinecite{rrnmnfkgbsm}] is that thermal shifts in the peak positions 
by factors of up to 3 are not accompanied by thermal broadening of the 
excitations over the same energy width. That the magnon modes remain 
coherent shows that their spectral--weight shifts are a true many--body 
effect governed by the overall band structure and by their hard--core 
statistics. These features are captured over the full range of temperatures 
(up to $T \sim J$) in the bond--operator framework with Eq.~(\ref{eehcbs}). 

Heuristically, a band--narrowing effect arises because, while the 
temperature does not alter the magnetic exchange interactions, thermal 
occupation of some dimer sites by triplets acts to block the motion of a 
propagating test triplet due to the local hard--core constraint (\ref{eboc}). 
This is reflected in the fact that the effective occupation function for 
mode $\alpha$ (\ref{eehcbs}) involves explicitly terms due to all three 
magnon branches, thus encoding a form of inter--branch exclusion. In the 
self--consistent solution of the mean--field equations, an increase 
in ${\tilde n} (\omega_{k,\alpha})$ with increasing temperature causes 
a drop in all of the quantities $\overline{s}$, $P$, and $Q$. The physical 
interpretation of this effect is a suppression both of on--dimer singlet 
condensation and in the coherence of interdimer fluctuation processes, 
which results in a narrowing of the magnon bands and consequent increase 
in the gap. While such an effect is also obtained with bosonic statistics, 
only the effective occupation function of Eq.~(\ref{eehcbs}) takes into 
account correctly the number of available hard--core boson states of the 
system.  

In this section, the qualitative physics of the two--chain spin ladder at 
finite fields and temperatures is illustrated using the isotropic coupling 
ratio, $J'/J = 1$. This value is not chosen for any special reason: it is 
already clear from the previous section that the bond--operator description 
is not quantitatively accurate beyond $J'/J = 0.5$, and it is not the case 
that some of the qualitative effects become stronger as the coupling ratio 
increases (indeed, as shown in Sec.~V, the converse is true for certain 
quantities). The choice is simply a good representative value for a ladder 
away from the strongly dimerized limit, and serves as an instructive 
counterpoint to the results presented in Sec.~V for $J'/J = 1/3.9$ (a 
value which is effectively in this limit). 

\begin{figure}[t]
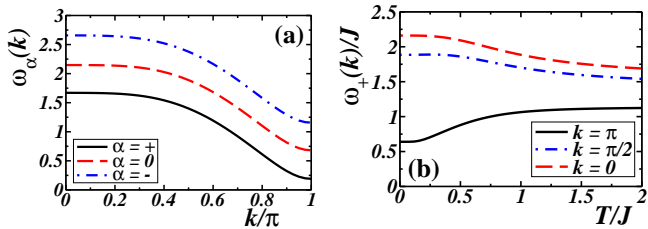

\includegraphics[width=4.1cm]{lbai.eps}\hspace{0.3cm}\includegraphics[width=4.1cm]{lgati0b.eps}
\caption{(color online) (a) Magnon bands $\omega_{\alpha} (k)$, shown for 
the isotropic ladder at $T/J = 0.1$ and in a field $h/J = 0.5$. (b) Thermal 
evolution of the band maximum, center, and minimum for a spin ladder with 
$J'/J = 1$ and $h = 0$. } 
\label{lf3}
\end{figure}

\subsection{Magnon dispersion}

Figure 3(a) serves first as a reminder of the magnon band structure: there 
are three branches, which all have a minimum at the antiferromagnetic wave 
vector $k = \pi$ and which are similar in shape and splitting. However, for 
a ladder beyond the strongly dimerized limit ($J' \ll J$), and at finite 
fields and temperatures, these branches are not symmetrical (in $k$) about 
a band center, they do not have exactly the same line shape, and their 
separation is not exactly the bare Zeeman splitting defined in Sec.~IIIA; 
all of these facts can be read from Eqs.~(\ref{ewp})--(\ref{ew}). 

\begin{figure}[t]
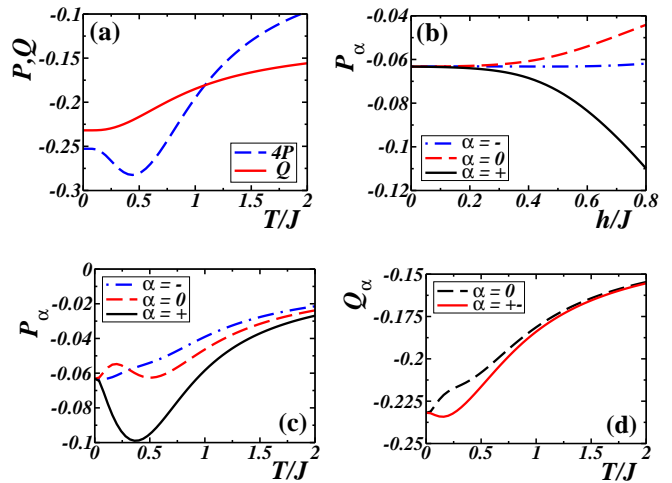

\includegraphics[width=4.0cm]{lpqati0b.eps}\hspace{0.3cm}\includegraphics[width=4.3cm]{lpaib.eps} \\
\phantom{.}
\includegraphics[width=4.0cm]{lpatib.eps}\hspace{0.3cm}\includegraphics[width=4.1cm]{lqatib.eps}
\caption{(color online) (a) Triplet correlations $P$ (blue, dashed) and $Q$ 
(red, solid) as functions of temperature for $J'/J = 1$ and $h = 0$. $P$ is 
shown multiplied by a factor of 4 to illustrate its relative thermal 
renormalization. (b) Triplet correlations $P_+$, $P_0$, and $P_-$ shown 
as functions of field at $T/J = 0.1$. (c) $P_+$, $P_0$, and $P_-$ and (d)
$Q$ and $Q_0$ as functions of temperature for $J'/J = 1$ and $h/J = 0.5$.} 
\label{lf4}
\end{figure}

Considering first the case with no magnetic field, Fig.~3(b) illustrates 
the thermal band--narrowing effect by showing for the isotropic ladder the 
$T$--dependence of the band maximum ($k = 0$), band minimum ($k = \pi$), 
and band center defined as the mode energy at $k = \pi/2$. The suppression 
of hopping begins when $T$ becomes similar to the gap, and the effect  
saturates when $T$ exceeds $J$. The form of the thermal renormalization, and 
of the consequent growth of the gap with temperature, is exactly analogous to 
the results obtained for TlCuCl$_3$ in Ref.~[\onlinecite{rrnmnfkgbsm}]. As 
noted above, $\omega_{k,\alpha}$ obtained from the bond--operator formalism 
is the peak position at finite temperture of the one--magnon spectral weight 
of the spin ladder, a quantity not yet available from any numerical 
approaches. However, some recent attempts to compute effective magnon 
spectra for quantum magnets at finite temperatures, specifically ED 
calculations for short, dimerized (gapped) $S = 1/2$ chains\cite{rml} 
and real--time DMRG calculations for gapless $S = 1/2$ Heisenberg 
chains,\cite{rbsw} suggest that further progress can be expected in 
this direction. 

The evolution of the zero--field triplet correlations $P$ and $Q$ 
corresponding to this band--narrowing is shown in Fig.~4(a). The two 
quantities have contrasting behavior: while finite temperatures act only 
to suppress pair--creation processes ({\it i.e.}~rendering these less 
coherent), the hopping correlations show an initial rise, or thermal 
enhancement of such processes, before thermal decoherence becomes the 
dominant effect. Here the terms ``coherence'' and ``decoherence'' are 
used in the sense of Sec.~III, the former to indicate the intrinsic 
correlations driven by quantum fluctuations (which depend on $J'/J$, 
Fig.~2) and the latter to indicate the suppression of these correlations 
by thermal fluctuations. 

At finite fields and temperatures, the full complexity of the system 
becomes visible, with effects appearing which are not present at finite 
$h$ or $T$ alone. Figures 4(c) and (d) show the thermal evolution of the 
triplet correlation functions at finite $h$: $P_+$ is strongly and $Q$ 
weakly enhanced at small fields, before thermal decoherence takes over; 
$P_-$ is generally anticorrelated to $P_+$, but has no mechanism for 
initial suppression corresponding to the enhancement of $P_+$; $P_0$ 
shows a non--monotonic evolution as a consequence of the competing 
effects of coupling to $P_+$, the field scale, and the two opposing 
forms of thermally driven behavior [Fig.~4(a)]; $Q_0$ shows the same 
weak anticorrelation to $Q$, followed by thermal decoherence [Fig.~4(d)]. 
The correlation functions vary somewhat predictably with $h$ when $T$ is 
fixed, as shown in Fig.~4(b), although it should be noted that the flat 
line is $P_-$, while $P_0$ is the function which drops more strongly as 
$P_+$ is enhanced. This result can be understood by considering the 
corresponding temperature in Fig.~4(c). 

It is important to note at this point that it is not particularly meaningful 
to discuss temperatures $T$ in excess of, or even on the order of, the rung 
coupling scale $J$. At these temperatures, the concept of the rung singlet 
and triplets as well--defined states of the two rung spins breaks down, 
and it is more appropriate to consider these spins separately. Even at 
temperatures considerably below $J$, magnon excitations become broad 
functions in the space of energy and wave vector: the complex question of 
this intrinsic line width in a gapped quantum magnet has been considered 
only at the semiclassical level,\cite{rds} and more microscopically in some 
limiting cases,\cite{re} but is not addressed within the bond--operator 
framework at the current level of approximation. Thus the data up to $T = 
2 J$ shown in Figs.~3 and 4 are to be taken only as an example of the physics 
encoded in the mean--field equations, and all of the figures to follow will 
be cut at $T = J$. For those concerning the critical properties of the ladder 
near the field--driven crossover to the Luttinger--liquid phase, further 
caveats will be necessary. 

\begin{figure}[t]
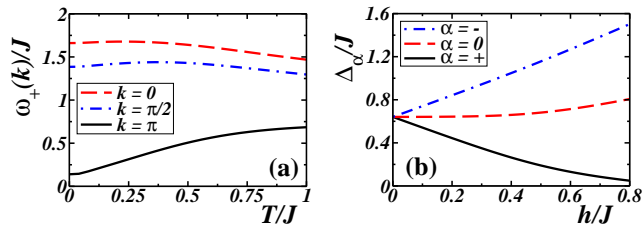

\includegraphics[width=4.0cm]{lgatib.eps}\hspace{0.3cm}\includegraphics[width=4.1cm]{lgaib.eps}
\caption{(color online) (a) Band--narrowing effects for the isotropic ladder 
at $h/J = 0.5$. (b) Gap--curvature effects for the band minimum of the three 
triplet branches, illustrated for the isotropic ladder with $T/J = 0.1$. }
\label{lf5}
\end{figure}

The thermal renormalization of the triplet band width at finite field 
is shown in Fig.~5(a). The effect of the field is to reduce further the 
symmetry of the band edges, causing an initial upward trend at the upper 
edge as thermal fluctuations allow the polarizing effect of the field 
to operate. Analogous to the effects visible in Figs.~3 and 4, this 
behavior is reversed at higher temperatures, as $T$ competes with $h$, 
resulting in an overall downward trend of the band center. The consequences 
for the mode gaps as a function of field at finite $T$ are shown in Fig.~5(b): 
here one may regard the 0--mode curvature as an indication of generic thermal 
effects (increased net energy), on top of which the additional contributions 
contained in the $P$ and $m$ terms in Eqs.~(\ref{ewp}) and (\ref{ewm}) act 
against this curvature for the upper--mode gap, giving a rather straight 
field--dependence, but an enhanced curvature away from gap closure of the 
lower mode. This reflects the basic physical phenomenon that the lower mode 
absorbs most of the thermally excited magnons at finite $T$ and $h$, and is 
thus the most strongly renormalized; the necessary deviation from Zeeman 
splitting is contained in the $P$ and $m$ terms. 

\begin{figure}[t]
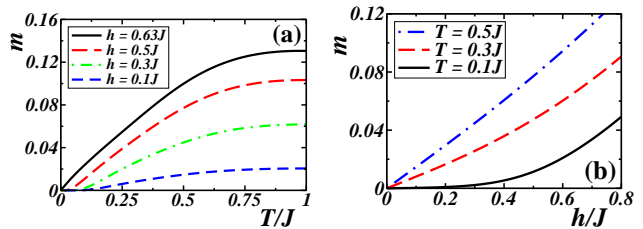

\includegraphics[width=4.0cm]{lmmatic.eps}\hspace{0.3cm}\includegraphics[width=4.0cm]{lmmaic.eps}
\caption{Magnetization $m$ as a function of $T$ for four different magnetic 
fields (a) and as a function of $h$ for three different temperatures (b).}
\label{lf6}
\end{figure}

\subsection{Thermodynamic properties} 

It is instructive to consider the magnetization $m$ to understand its 
evolution with field and temperature, and hence how it affects the magnon 
dispersion branches. For a fixed field as a function of temperature 
[Fig.~6(a)], $m$ is non--monotonic, first rising as a consequence of the 
thermally excited states which become available to react to the field, but 
becoming maximal as $T$ approaches $J$. The low--temperature rise is clearly 
of activated form, the gap diminishing as the field is raised. Only close to 
the critical field which closes the gap ($h/J = 0.64$ in the isotropic ladder) 
does the exponential form turn over to a power law, which in 1D is expected 
to be a square root as a function of temperature at very low $T$. The value 
of $m$ at the maximum is clearly an almost linear function of the field which 
is polarizing the available triplet states. The high--$T$ behavior, found 
also in extensive TMRG studies of $m(h,T)$,\cite{rwy} sets in when $T$ 
competes with the interaction energy scales of the system, suppressing not 
only triplet propagation and the polarized moment, but also dimer singlet 
formation itself. 

Figure 6(b) shows $m$ as a function of $h$ for different temperatures. In 
general, both the number of states becoming available and the field which 
is able to polarize them give contributions to $m$. At low $T$ the response
is activated and the dominant contribution is due to the field. At high $T$, 
for which $T/J = 0.5$ is already sufficient, $m$ becomes an entirely linear 
function of $h$: here the temperature ensures a large number of available 
states, and these are polarized by whatever field is applied. This physics, 
also illustrated below in Figs.~9(e) and (f), constitutes a regime in which 
the gap renormalization [of the type shown in Fig.~5(b)] is also completely 
linear. This results in a ``pseudo--Zeeman'' behavior of the mode minima in 
an applied field, where the gaps evolve linearly but with a slope reduced 
from the conventional, zero--temperature value by the $P_+ - P_-$ and $m$ 
terms in Eqs.~(\ref{ewp}) and (\ref{ewm}). It is possible that this behavior 
has been misinterpreted in some previous analyses of low--dimensional quantum 
magnets, for example as a $T$--dependent $g$--factor. 

At low temperatures one finds that $m \simeq {\textstyle \frac{1}{2}} (P_-
 - P_+)$, and thus that the two contributions to non--Zeeman behavior of 
the gap are essentially equal. This value of $m$ can be regarded as the 
quantum--fluctuation contribution to the magnetization. As $T$ is increased, 
$m$ increases compared to $|P_+ - P_-|$, until at high temperatures it is 
exponentially greater: while $m$ is enhanced by thermal population effects, 
there is no such contribution to $|P_+ - P_-|$, which indeed is suppressed 
at high temperatures [Fig.~4(c)]. 

In dimensions of 2 and higher, it is of great interest to investigate the 
temperature--dependence of the critical field $h_c$, the field which closes 
the excitation gap of the quantum spin system, driving a quantum phase 
transition. The quantity $h_c(T) - h_c(0)$ encapsulates the critical 
properties of the theory, and has been analyzed in detail for the 
higher--dimensional cases, where the transition is in the Bose--Einstein 
universality class.\cite{rgrt} In this case, the transition is from a 
(singlet--based) quantum disordered state to an ordered magnetic 
one, the order parameter being the transverse (staggered) moment. The 
low--temperature behavior of the phase--transition line should have the 
scaling form $h_c(T) - h_c(0) = A T^{d/2}$, where A is a constant, $d$ is 
the spatial dimensionality of the system, and the denominator 2 in the 
exponent enters from the quadratic dispersion of the free hard--core 
bosons ($\omega \propto k^2$) at the transition. 

\begin{figure}[t]
\includegraphics[width=8.7cm]{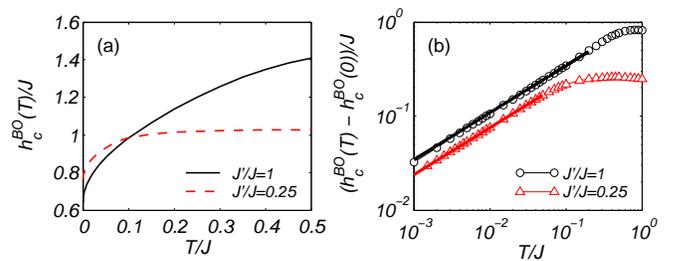}
\caption{(a) Critical magnetic field $h_c^{\rm BO}$ required to close the 
effective spin gap of ladders with coupling ratios $J'/J = 1$ and 0.25 at 
finite temperatures. (b) $h_c^{\rm BO} (T) - h_c^{\rm BO} (0)$ on logarithmic 
axes to illustrate square--root scaling (solid lines). }
\label{lf7}
\end{figure}

In a 1D system, there is no ordered phase or order parameter, and indeed 
no transition: $h_c (T)$ marks a crossover to the Luttinger--liquid 
regime.\cite{rsss,rcrct} Under these circumstances, the definition 
of a quantity $h_c (T)$ may not be unique. The crossover has nevertheless 
been found to have a very clear signature both in experimental measurements 
of the thermodynamic properties of (C$_5$H$_{12}$N)$_2$CuBr$_4$ and in 
numerical studies of the magnetization $m(h,T)$. In measurements of the 
magnetocaloric effect,\cite{rcrct} its location was deduced from the minima 
and maxima in $m(T)$ through the condition $\partial m/ \partial T |_h = 0$, 
while in TMRG calculations\cite{rwy} it was taken from the maximum in 
$\partial m/ \partial h$. The quantity $h_c^{\rm BO} (T)$ investigated in 
the bond--operator description is that value of the field where the maximum 
of the finite--temperature, one--magnon spectral function is moved to zero 
frequency, and therefore does not necessarily have a direct relation to 
the crossover field (which is dictated by the lowest level in the 
excitation spectrum): it is rather a statement about the reaction of 
the spectral--weight function to thermal flucutations, which may be very 
different in 1D from the higher--dimensional cases. For practical purposes, 
$h_c^{\rm BO} (T)$ marks the limit of applicability of the bond--operator 
formalism at the level of Sec.~II, which cannot be used for the gapless 
phase without extensive modification. At $T = 0$, the quantity $h_c = 
h_c^{\rm BO} (0)$ is the same crossover field as that obtained from all 
other definitions. 

The critical field $h_c^{\rm BO} (T)$ is shown in Fig.~7(a) for the 
isotropic ladder, and by way of comparison also for a ladder with coupling 
ratio $J'/J = 1/4$. At low temperatures, both curves have a very rapid 
increase with $T$: the logarithmic fit in Fig.~7(b) reveals this singular 
behavior to have precisely the square--root form obtained from the analysis 
above with $d = 1$. Thus the effect of thermal fluctuations on the 
low--temperature spectral function is indeed dramatic. Beyond the 
critical regime, $h_c^{\rm BO} (T)$ becomes flatter at higher temperatures, 
meaning that the maximum of the excitation spectrum becomes rather 
temperature--insensitive. A slight decrease is visible for $J'/J = 1/4$ as 
the system approaches the ``breakdown'' regime $T \sim J$, where the theory 
contains diminishing energy scales [visible in the magnon bands (Figs.~3 and 
5) and in the magnetization (Fig.~6)]. Returning to the critical regime, 
while it may be taken to exist over a temperature range of order $0.2J$ in 
the isotropic ladder, it is clearly correspondingly narrower for smaller 
coupling ratios ($T/J \sim 0.05$ at $J'/J = 1/4$). This narrow 1D critical 
regime is precisely that part cut off by 3D coupling even in very weakly 
interacting, quasi--1D ladder systems such as (C$_5$H$_{12}$N)$_2$CuBr$_4$. 

A further thermodynamic quantity which can be computed in this framework 
is the magnetic specific heat. Obtained from the second temperature 
derivative of the effective hard--core boson free energy, this can be 
expressed in the form 
\begin{equation}
C_m (T) \! = \!\! \sum_{k,\alpha} \! \frac{\beta^2 \omega_{k,\alpha}^2 
e^{- \beta \omega_{k,\alpha}}}{1 + \sum_\alpha \! z_\alpha (\beta)} \! - \! 
\left( \frac{\beta \omega_{k,\alpha} e^{- \beta \omega_{k,\alpha}}}{1
 + \sum_\alpha \! z_\alpha (\beta)} \right)^2 \!\!\! . 
\label{ecmt}
\end{equation}
Consideration of the magnetic specific heat is deferred to Sec.~V, where 
it is discussed in the context of a direct experimental comparison.

\section{$\rm \bf (C_5H_{12}N)_2CuBr_4$} 

Having understood in full the physics contained in the complete mean--field 
equations, one may turn to the nearly ideal spin--ladder material 
(C$_5$H$_{12}$N)$_2$CuBr$_4$. Fits to the magnon bands, of the type 
shown in Fig.~3(a), were performed in Ref.~[\onlinecite{rbtins}] and 
will not be repeated here. In the bond--operator framework they give 
$J = 13.07$ K, $J' = 3.35$ K, and hence a coupling ratio within the spin 
ladders of $J'/J = 1/3.9$.\cite{rbtins} In this strongly dimerized regime 
(sometimes referred to as the ``strong--coupling'' regime), the bond--operator 
description is extremely accurate (Sec.~III). 

The real material has a number of complications which may affect the 
accuracy of a detailed comparison between this theoretical description and 
experimental measurement on levels of up to 5\%. For the present purposes, 
the most important is the anisotropy in the $g$--factor,\cite{rpsw,rcea} which 
enters the comparisons between thermodynamic measurements (most performed 
with ${\vec H} \parallel {\hat a}$) and INS dispersion relations (${\vec H} 
\parallel {\hat b}$). The value chosen here to reproduce $H_c = 6.99$ T, 
$g = 2.12$, is close to the average in the $(ab)$--plane of the system. 
Among the other materials questions, one is the weak interladder coupling 
mentioned in Sec.~I, which is approximately 3\% of $J$;\cite{rbtins} 
another is a detectable magnetostriction effect (changes of $J$ and $J'$ with 
applied field) of order 1\%, which becomes relevant when comparing the low-- 
and high--field gapped regimes;\cite{rbtins} a third is a spin anisotropy, or 
a breaking of SU(2) spin symmetry of as--yet undetermined origin, which a 
recent ESR investigation\cite{rcea} has estimated as high as 5\% of $J$. 
While a spin anisotropy can be included in the current framework when enough 
information becomes available, the high--field regime is not the focus of 
the current manuscript and the 3D nature of the material is not of prime 
importance while the system has a robust spin gap, meaning at all fields 
below where $h$ enters the critical regime close to $h_c$. 

In this section, results are presented only for 1D ladders with coupling 
ratio $J'/J = 1/3.9$. Fig.~8 shows the triplet correlations $P$ and 
$Q$ which assist a qualitative understanding of the ladders in 
(C$_5$H$_{12}$N)$_2$CuBr$_4$. Figure 9 shows the magnetization and 
one--magnon band structure of these ladders at finite fields and 
temperatures, giving direct experimental predictions. Figure 10 
considers the thermodynamics of these ladders through the magnetic 
specific heat, including a direct experimental comparison and 
quantitative reproduction of published data for this quantity. 

\begin{figure}[t]
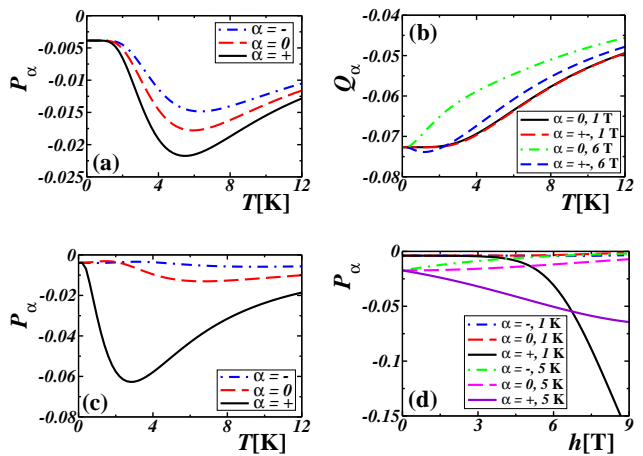

\includegraphics[width=4.0cm]{lpatm1b.eps}\hspace{0.3cm}\includegraphics[width=4.0cm]{lqatm16.eps} \\ 
\phantom{.}
\includegraphics[width=4.0cm]{lpatm6b.eps}\hspace{0.3cm}\includegraphics[width=4.0cm]{lpam15.eps}
\caption{(color online) Triplet correlation functions for the ladders of 
(C$_5$H$_{12}$N)$_2$CuBr$_4$ at different temperatures and applied magnetic 
fields. (a) $P_{\alpha}$ for $\alpha = +,0,-$ as functions of temperature 
at a field of 1 T. (b) $Q_0$ and $Q_{+-} = Q$ as functions of temperature 
at fields of 1 T and 6 T. (c) $P_{\alpha}$ at a field of 6 T. (d) $P_\alpha$ 
as a function of field for temperatures of 1 K and 5 K. }
\label{lf8}
\end{figure}

Beginning with the triplet correlations, the quantities $P_{\alpha}$ and 
$Q_{\alpha}$ are shown in Fig.~8 for ladders with $J'/J = 1/3.9$. It is 
clear in Fig.~8(a) that the general enhancement of hopping correlations 
by a finite temperature is large in strongly dimerized ladders, and that 
the scale for thermal decoherence to dominate the correlation physics of 
(C$_5$H$_{12}$N)$_2$CuBr$_4$ is $T \sim 10$ K. The same quantity shown not 
at low field but close to the critical point [Fig.~8(c)] emphasizes the 
very strong effect which the field can have on the enhancement of $P_+$ 
in this regime of $J'/J$ [note the different scales in panels (a) and (c)]. 
The triplet pair--creation correlation function $Q_\alpha$, however, depends 
strongly on neither field nor temperature, being sensitive only to the 
thermal decoherence effect [Fig.~8(b)]. As a function of the magnetic 
field [Fig.~8(d)], splitting of the hopping correlations in the three 
branches requires significant field strengths at low temperatures, where 
thermal occupations are activated, but the effect is nevertheless very 
strong on approaching the critical region. By contrast, at higher temperatures 
where states in all branches are available, all applied fields are capable 
of splitting the branches but the effect is rather modest. 

\begin{figure}[t]
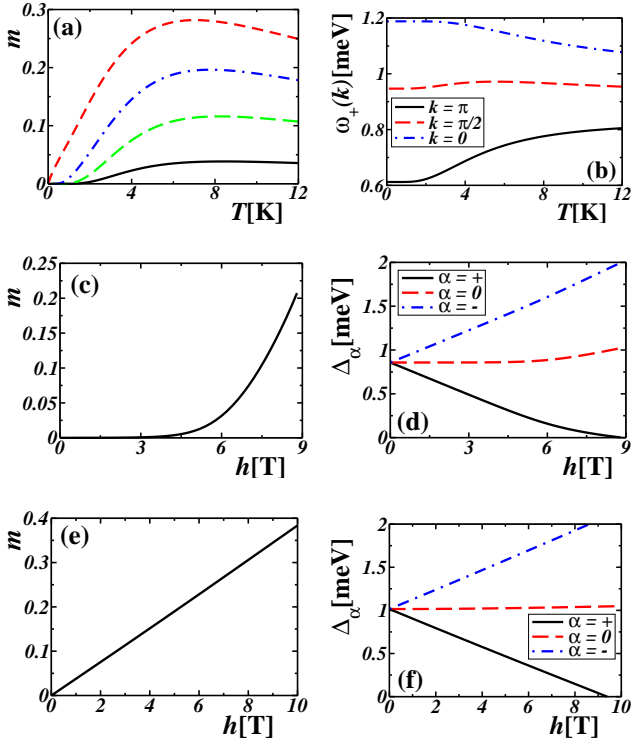

\includegraphics[width=4.0cm]{lmmatm.eps}\hspace{0.3cm}\includegraphics[width=4.0cm]{lgatm2.eps} \\ 
\phantom{.}
\includegraphics[width=4.0cm]{lmmam1b.eps}\hspace{0.3cm}\includegraphics[width=4.0cm]{lgam1b.eps} \\ 
\phantom{.}
\includegraphics[width=4.0cm]{lmmamb.eps}\hspace{0.3cm}\includegraphics[width=4.0cm]{lgamb.eps}
\caption{(color online) Calculations for (C$_5$H$_{12}$N)$_2$CuBr$_4$ ladders 
at finite temperature and magnetic field: as functions of temperature, (a) 
magnetization $m$ for fields of 1 T (black, solid line), 3 T (green, 
long--dashed line, 5 T (blue, dot--dashed line), and 6.99 T (red, dashed 
line) and (b) magnon band width at a field of 2 T; as functions of the 
applied magnetic field, magnetization $m$ (c,e) and gaps $\Delta_{\alpha}$ 
(d,f) at temperatures of 1 K (c,d) and 10 K (e,f). }
\label{lf9}
\end{figure}

Figure 9 turns to some of the observables of the ladder system. The 
magnetization as a function of temperature over a range of fixed magnetic 
fields is shown in Fig.~9(a): initial activated behavior dictated by the 
spin gap leads to a peak around $T \sim J/2$ and then to the slow decrease 
indicative of the loss of coherence in the magnetic sector. These features 
are qualitatively similar to those of the magnetic specific heat, shown in 
Fig.~10, but differ in magnitude because $m$ increases as the system 
approaches the critical field [Fig.~6(a)]. Figure 9(b) shows the thermal 
band--narrowing effect of Figs.~3(b) and 5(a), illustrated for a magnetic 
field of 2 T. This phenomenon was explained in Sec.~IV, and INS measurements 
are expected shortly. Data for the magnetization $m$ and for the gap 
$\Delta_{\alpha}$ of each mode as functions of the applied field, at 
temperatures equivalent to 1 K and 10 K, are shown in Figs.~9(c)--(f). 
The magnetization at 1 K [Fig.~9(c)] shows the activated behavior visible 
at lower temperatures in Fig.~6(b), and the gap of the low--lying mode 
[Fig.~9(d)] has a significant curvature of the type shown in Fig.~5(b). 
By contrast, the 10 K data are well in the high--temperature limit 
discussed in Sec.~IV, where the ready availability of thermally 
excited states leads to a linear magnetization [Fig.~9(e)] and to 
the pseudo--Zeeman form of the mode gaps as functions of temperature 
[Fig.~9(f)]. It should be stressed, however, that in experiment it is 
somewhat unlikely that clearly defined magnon excitations would still 
be visible at temperatures as high as 10 K, due to extensive thermal 
broadening of the response function. If one considers temperatures of 
approximately 5 K, where a definite magnon mode may still be detectable, 
then these data are rather similar to the forms shown in Figs.~9(e) and 
(f), albeit with some discernible curvature remaining in $m$ and $\Delta_+$.

The most directly measurable and perhaps informative single thermodynamic 
quantity is the magnetic specific heat. Detailed measurements performed on 
(C$_5$H$_{12}$N)$_2$CuBr$_4$ over the full range of temperatures, and of 
applied magnetic fields in both the gapped (quantum disordered) and gapless 
(Luttinger--liquid) phases, are presented in Ref.~[\onlinecite{rcrct}]. 
The experimental results were compared with somewhat numerically demanding 
ED and DMRG studies of ladders with the same coupling ratio, and also to a 
bond--operator approximation in the gapped regime and a Bethe--Ansatz 
approximation in the gapless one. The same behavior of the magnetic specific 
heat as a function of field and temperature has also been demonstrated in 
TMRG studies of ladder thermodynamics,\cite{rwy} but has not been computed 
explicitly for the parameters of (C$_5$H$_{12}$N)$_2$CuBr$_4$. 

\begin{figure}[t]
\includegraphics[width=8.7cm]{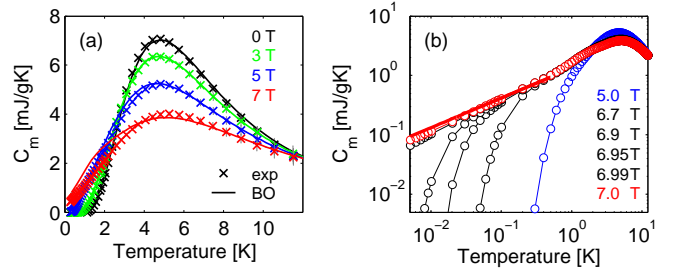}
\caption{(color online) (a) Magnetic specific heat of 
(C$_5$H$_{12}$N)$_2$CuBr$_4$ at magnetic fields of 0 T (black, uppermost 
peak), 3 T (green, second peak), 5 T (blue, third peak) and 7 T (red, 
lowest peak). Solid lines are given by the complete bond--operator theory, 
while crosses show experimental data presented in Ref.~[\onlinecite{rcrct}]. 
(b) Specific--heat data from bond--operator theory, shown on logarithmic axes 
to illustrate the approach to a low--$T$ square--root form as $h \rightarrow 
h_c$. The red, solid line marks a slope of 1/2.} 
\label{lf10}
\end{figure}

The magnetic specific heat given by the complete bond--operator approach 
is shown in Fig.~10(a) for fields corresponding to 0 T, 3 T, 5 T, and 7 
T, along with experimental measurements made at the same fields.\cite{rcrct} 
In the regime with a robust spin gap, the specific heat shows exponential 
activation at low temperatures, followed by a peaking at $T \sim J/2$. The 
activation gap decreases as the field is raised, resulting in more weight 
at lower temperatures and a drop in the peak height. By contrast, the 
critical field which closes the triplet gap in a single 
(C$_5$H$_{12}$N)$_2$CuBr$_4$ ladder is 6.99 T, and hence the system shows 
gapless behavior at 7 T; note that while the bond--operator description is 
no longer valid at $h > h_c$ for $T = 0$, the rapid rise of $h_c^{\rm BO}$ 
with temperature shown in Fig.~7 ensures that the gapless nature of the 
excitations becomes a numerical problem only at temperatures two orders 
of magnitude below those shown in Fig.~10(a). As the field approaches 
$h_c$, the peak height at $T \sim J/2$ continues to fall; both this 
peak and the high--$T$ decrease are features sensitive not to the gap but 
to the center and upper edges of the excitation band. 

The low--$T$ form of the specific heat in the gapless regime is a power 
law, and the power in a spin ladder at the critical field is 1/2. Figure 
10(b) shows specific--heat data obtained as $h$ approaches $h_c$ from 
below: clearly a square--root dependence is obtained, but only for fields 
extremely close to the critical value ({\it cf.}~Ref.~[\onlinecite{rlea}]). 
That even differences of 0.01 T can have a significant impact on the form 
of thermodynamic quantities indicates at the theoretical level the 
possibility that the quantum critical regime may be extremely narrow, 
and at the experimental level the need for exquisite accuracy in the 
measurement of critical properties. 

The data computed from Eqs.~(\ref{emfem})--(\ref{emfemm}), represented as 
lines in Fig.~10(a), is in perfect quantitative agreement with essentially 
every feature of the 0 and 3 T experimental data, and shows only the most 
minor deviations at 5 T. There is no separate scale factor for each curve, 
{\it i.e.}~as already stressed above there are no free parameters in the 
theory. Only on the critical line, at the very limit of applicability of 
the bond--operator framework, are some small but systematic discrepancies 
noticeable: specifically, the complete bond--operator result gives a peak 
height (around 5 K) which is marginally too low, and this missing weight 
appears to be present around temperatures of 2 K. However, the results of 
Fig.~10(b) indicate that extreme caution is required before ascribing this 
to a failure of the bond--operator description, as the true proximity of 
the experimental measurements to the critical field is accurate perhaps 
only to within 0.05 T. 

The bond--operator approximation applied in Ref.~[\onlinecite{rcrct}] used 
the same effective magnon statistics as here, but not the $P$, $Q$, and $m$ 
terms; the same qualitative form of error was found when $h \rightarrow h_c$ 
as in Fig.~10(a), but it was very much larger (of order 15\% in the peak 
height). Figure 10(a) makes clear that the self--consistent inclusion of 
these terms results in a very significant correction of this error. Thus 
the complete bond--operator framework is highly accurate even at the very 
limit of its regime of applicability and even in one dimension. It was noted 
in Ref.~[\onlinecite{rcrct}] that some degree of error may be introduced by 
the treatment of the hard--core constraint (\ref{eboc}): while the total 
number of states is correct, global application of the constraint may result 
in some of these states not being penalized sufficiently for the fact that 
they violate the constraint at the local level. Extra weight would then 
be expected not in the peak but at lower energies, perhaps resulting in 
the apparent persistence of the low--$T$ critical (power--law) regime to 
artificially high temperatures. However, the results in Fig.~10 show that 
these errors are at the percent level even for $h = h_c$, and that the 
complete bond--operator theory is capable of a fully quantititively 
accurate description of all features of the gapped, strongly dimerized 
spin ladder. 

A more explicit summary statement on the accuracy of the complete 
bond--operator treatment for the two--chain spin ladder, drawn from 
all of the results above, is the following: it is extremely high, meaning 
quantitative agreement within a few percent when compared to numerical 
results (particularly DMRG for the spectrum and TMRG for thermodynamics) 
for the strongly dimerized regime $J \gg J'$. This level of agreement, 
which is achieved with no adjustable parameters, extends to $J' \simeq J/2$ 
and applies to all physical observables computed here (spin correlations, 
magnon spectra, thermodynamic quantities). Quantitative deviations from 
percent levels of accuracy become noticeable as $J'/J$ is further increased 
towards the isotropic case, a result not at all surprising in view of the 
expected regime of applicability of the bond--operator method, and the type 
of quantum fluctuations which it is designed to capture. In the isotropic 
ladder, where the qualitative physics of the system is nevertheless very 
well described within the bond--operator approach (hence its use for 
illustrative purposes in Sec.~IV), the level of quantitative error as 
judged from the interdimer spin correlations and the zero--field spin 
gap is of order 20\%.

\section{Summary}

A complete version of the bond--operator theory for the spin ladder has 
been developed, which captures all of the relevant physics of the quantum 
disordered system, at all temperatures and all magnetic fields up to the 
closure of the spin gap. [At this point the magnon excitations fractionalize 
into spinons and a Luttinger--liquid description is appropriate.] At the 
heart of the finite--temperature framework is a global ansatz for the 
effective exclusion statistics of the magnon excitations, which are 
hard--core bosons. Among the phenomena described with quantitative 
accuracy and with no free parameters are the spin correlations, the 
thermal evolution of the magnon bands, and their thermodynamic properties. 
Extensive comparison becomes possible with existing data for the spin--ladder 
material (C$_5$H$_{12}$N)$_2$CuBr$_4$, and predictions are made for 
forthcoming experiments which will investigate the thermal renormalization 
of the magnon excitations. The same treatment is also applicable to calculate 
finite--temperature properties in the quantum disordered phases of other 
coupled--dimer systems of any dimensionality.

\acknowledgments 
 
The authors are grateful to T. Giamarchi, M. Sigrist, and G. S. Uhrig for 
helpful discussions, to E. Gull for computing assistance, and to K. P. Schmidt 
and G. S. Uhrig for provision of the CUT data shown in Figs.~2(c) and (f). 
The Pauli Center for Theoretical Studies at ETH Zurich and the Department 
of Physics of the University of Fribourg extended their generous hospitality 
to BN during the completion of the manuscript. This work was supported by 
the National Science Foundation of China under Grant No.~10874244, by 
Chinese National Basic Research Project No.~2007CB925001, and by the Royal 
Society.

\end{document}